# Cold atmospheric plasma activated deionized water using helium, argon, and nitrogen as feeding gas for cancer therapy

Zhitong Chen[1,2]

[1]National Innovation Center for Advanced Medical Devices, Shenzhen 518000, China
[2]Department of Mechanical and Aerospace Engineering, The George Washington University, Washington, DC 20052, USA

Email: zt.chen@nmed.org.cn

**Abstract:** Plasma activated medium, containing reactive oxygen species (ROS) and reactive nitrogen species (RNS), has been used for cancer treatment as an indirect treatment method. Here we report cold atmospheric plasma (CAP) activated deionized (DI) water using helium (He), argon (Ar), and nitrogen ($N_2$) as feeding gas for the cancer therapy. Basic characteristics of CAP activated DI water with different feeding gases were diagnosed and the effects of three solutions on the breast cancer cells and gastric cancer cells were also recorded. Our findings show that DI water treated by CAP with He feeding gas has a much stronger effect on apoptosis in precultured breast cancer cells and gastric cancer cells. These results are attributed to the higher concentration of ROS generated in CAP-activated DI water using He as feeding gas.

CAP has been utilized in various traditional fields and is being gradually developed into more fields due to its unique nature.[1] It's known for the generation of reactive species (ROS and RNS), charged particles, electronically excited atoms, etc.[2-4] Plasma containing plenty of composition leads to itself a broad range of applications such as environment, materials, agriculture, semiconductor, medicine, and aerospace.[5-13] Among all these fields, plasma medicine is an innovative and emerging interdisciplinary research field that combines plasma physics, chemistry, life science, and clinical medicine.[14] As the temperature of CAP is close or even lower than room temperature, CAP seldom causes thermal damage to the human body and biological tissues. Due to the benefits from above, CAP is attracting increasing attention and has been used in sterilization, dental treatment, cosmetology, hemostasis, anti-inflammation, wound healing, and skin disease treatment.[15-18]

CAP has shown successful applications in cancer therapy both *in vivo* and *in vitro*, including oral cancer, lung carcinoma, throat cancer, breast cancer, neuroblastoma, hepatocellular carcinoma, melanoma, skin carcinoma, pancreatic carcinoma, colon carcinoma, cervical carcinoma, and bladder carcinoma.[19-28] Fridman et al reported the anti-cancer effect of CAP in 2007, promoting plasma medicine to a new level.[29] CAP tends to inhibit the growth of cancer cells rather than homologous normal cells in dozens of cell lines.[30,31] CAP treatment can be directly applied to the tumor/tissues/cells. For example, Li e*t al* employed the surface dielectric barrier discharge (S-DBD) to treat the human hepatocellular carcinoma cells.[32] In order to expand the application of CAP, plasma can be delivered through a tiny tube for cancer therapy due to the inaccessibility of conventional CAP devices. Chen *et al*. delivered CAP through micro-sized tube to glioblastoma, which effectively prevented glioblastoma growth both *in vitro* and *in vivo*.[33,34] In addition, plasma-activated medium/liquid can be prepared in advance and stored until use for cancer therapy.[35,36]

In this work, we designed a plasma device submerged in DI water using He, Ar, and $N_2$ as feeding gas (Fig. 1a). The voltage and current of plasma-activated DI water were measured with a Tektronix TDS 2024B Oscilloscope. The spectra of plasma-activated DI water were characterized by UV-visible-NIR Optical Emission Spectroscopy. The temperature of the plasma solution was measured with FLIR Systems Thermal Imaging. The concentrations of ROS and RNS in DI water were determined by using a Fluorimetric Hydrogen Peroxide Assay Kit (Sigma-Aldrich, MO), and the Griess Reagent System (Promega, WI). The cell viability of the human gastric cancer cell line (NCl-N87) and breast cancer cell lines (MDA-MB-231 and MCF-7) were monitored with the Cell Counting Kit 8 assay (Dojindo Molecular Technologies, MD).

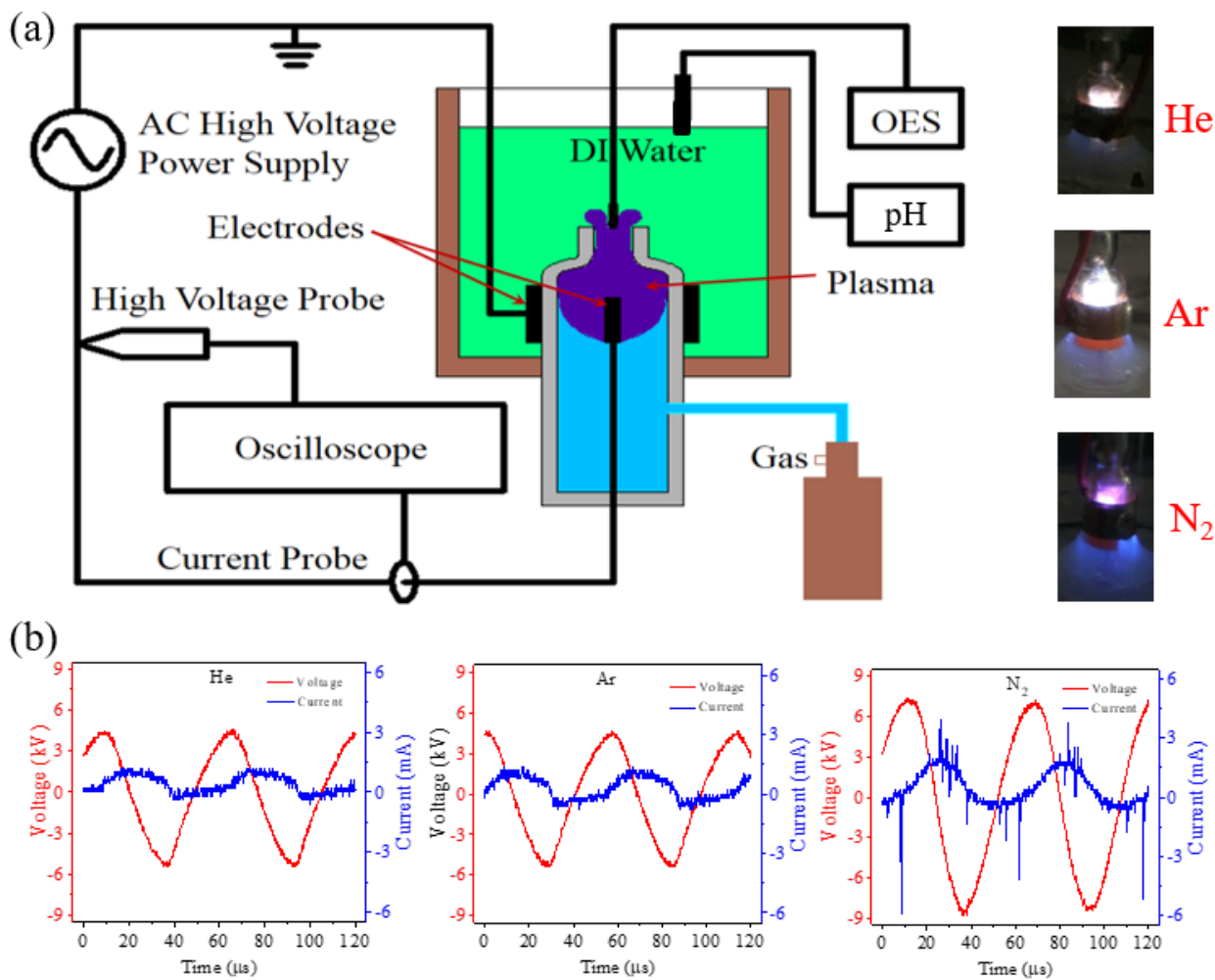

**Fig. 1**. (a) Schematic diagram of CAP device treated DI water, and images of plasma-treated DI water with He, Ar, and $N_2$ as feeding gas. (b) Discharge voltage and current of plasma discharge during treating DI water, using He, Ar, and $N_2$ as feeding gas.

Fig. 1a shows the CAP device submerged in DI water, which consists of 2 electrodes assembled with a central powered electrode and a grounded outer electrode wrapped around the outside of the quartz tube. The electrodes were connected to the high voltage. Images of plasma-activated DI water using $N_2$, He, and Ar as carrier gases are shown in Fig.1a. The flow rate for each feeding gas is 1.0 L/min. Fig. 1b demonstrates the voltage and current of plasma-activated

DI water using $N_2$, He, and Ar as carrier gases. The peak-peak voltage, as well as average current, of plasma-activated DI water using $N_2$ is much higher than Ar and He.

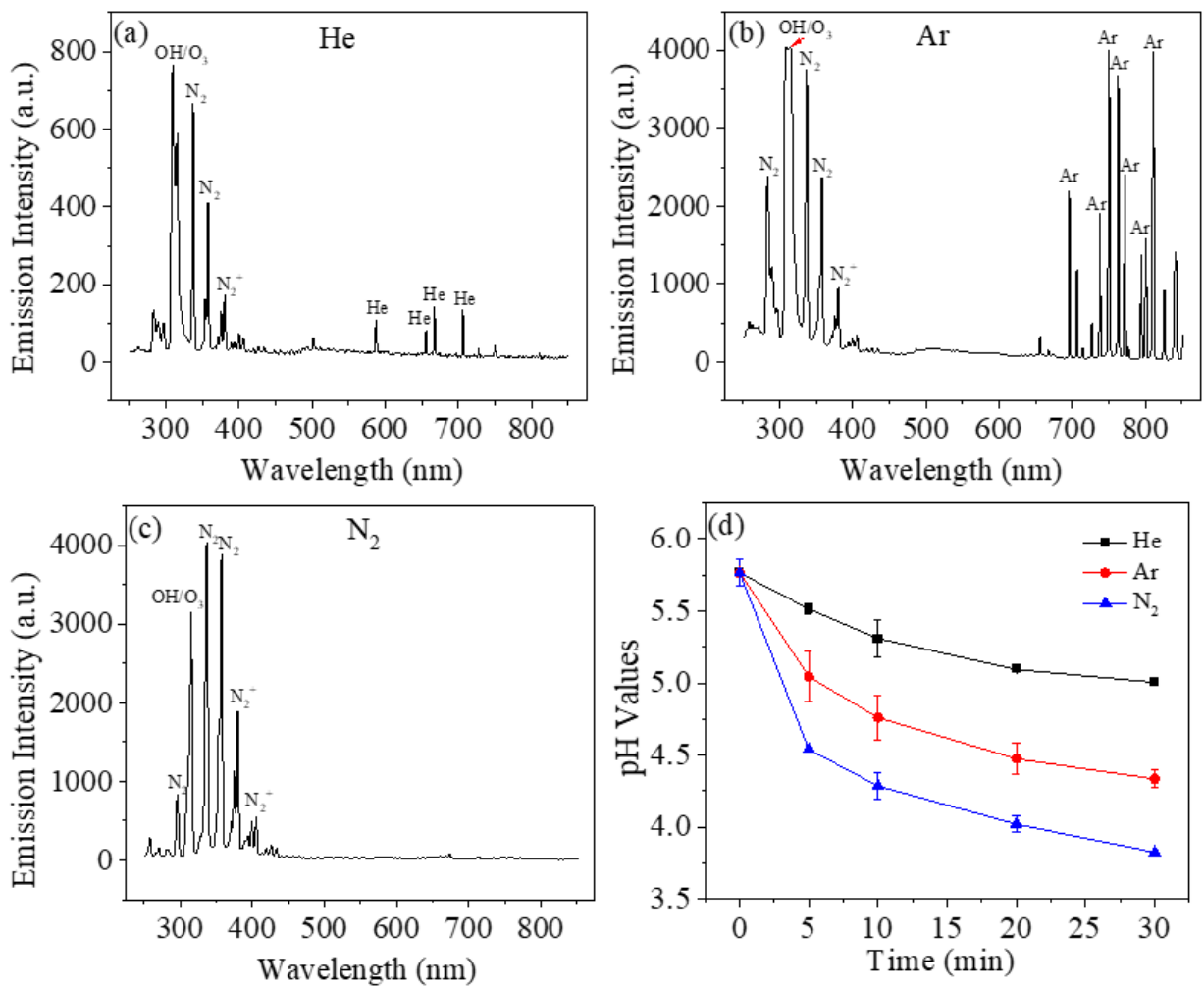

**Fig. 2**. The optical emission spectrum of plasma using He (a), Ar (b), and $N_2$ (c) as feeding gas from the CAP device submerged in DI water in the 250-800 nm wavelength range. (d) pH changes of plasma solutions using He, Ar, and $N_2$ as feeding gas based on treatment time.

Fig. 2 shows the optical emission spectrum of plasma-activated DI water using three feeding gases and the pH values of plasma-activated DI water using three gases. The reactive species generated during plasma-activated DI water using He, Ar, and $N_2$ are shown in Fig. 2a, Fig. 2b, and Fig. 2c, respectively. The identification of emission lines and bands was performed according to reference.[37] Using $N_2$ as a feeding gas leads to the formation of $N_2$ second-positive system (C3Πu-B3Πg) with its peaks at 316, 337, and 358 nm, which is higher than Ar and He, especially higher than He. For Ar, the intensity of OH peak is higher than peaks of $N_2$ second-positive system, similar phenomena were also observed in He plasma-activated DI water. Naturally, Ar and He lines in the range of 600 to 800 nm were observed. Fig. 2d shows pH values of plasma-activated DI water using three feeding gases after 5 min, 10 min, 15 min, 20 min, 25 min, and 30 min. The pH values demonstrate time-dependent behavior. At the same treatment time, the pH value of $N_2$ plasma is the highest, while the pH value of He plasma is the lowest. It is reasonable because $N_2$ plasma generates the highest intensity of RNS, while it is the lowest for He plasma.

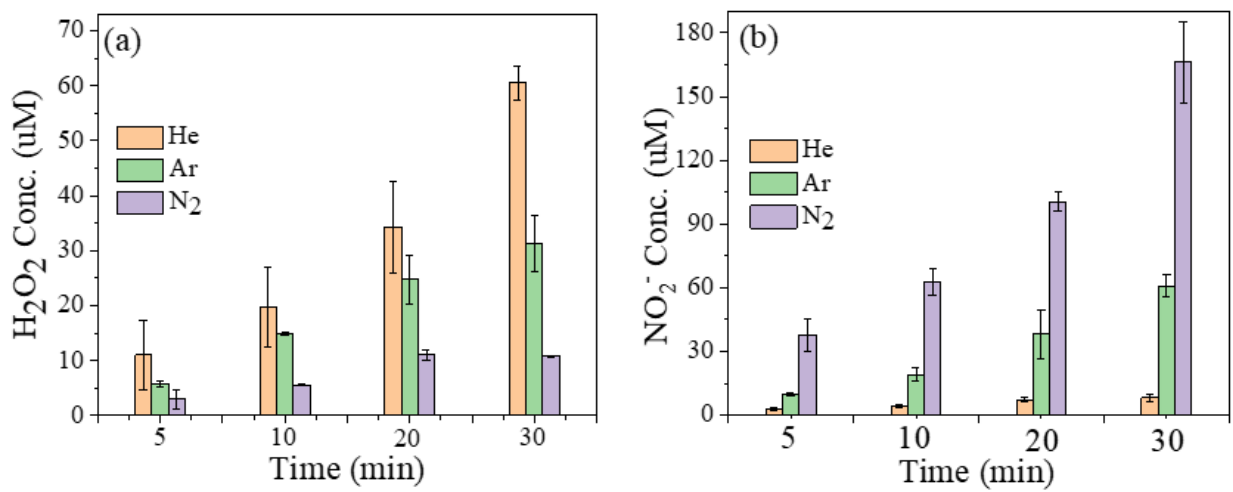

**Fig. 3**. $H_2O_2$ concentrations (a) and $NO_2^-$ concentrations (b) of plasma-treated water using He, Ar, and $N_2$ as feeding gas based on treatment time.

Plasma-activated DI water produces ROS and RNS such as the hydroxyl radical, superoxide, singlet oxygen, and nitric oxide.[38] In solution, these relatively short-lived ROS/RNS converts to relatively long-lived species such as hydrogen peroxide ($H_2O_2$), nitrite ($NO_x$), and other uncertain species. $H_2O_2$ and $NO_x$ are known to induce cell proliferation as well as cell death. $H_2O_2$ is known to induce both apoptosis and necrosis, while $NO_x$ can induce cell death via DNA double-strands break.[39,40] Fig. 3 shows the $H_2O_2$ and $NO_2^-$ concentrations of plasma-activated DI water using three feeding gases after 5 min, 10 min, 20 min, and 30 min. For $H_2O_2$ concentration, He plasma is the highest, while $N_2$ plasma is the lowest at the same treatment time. For $NO_2^-$ concentration, $N_2$ plasma is the highest, while He plasma is the lowest at the same treatment time. Recall that the relatively highest OH peak was observed in Ar plasma (as shown in Fig 2), however, its concentration of $H_2O_2$ is much lower than He plasma. $H_2O_2$ formation comes from •OH reaction (•OH + •OH → $H_2O_2$).[41] For $NO_2^-$ formation includes following reactions: 1) $N_2$ + e → 2N + e; 2) N + $O_2$ → NO + O; 3) 4NO + $O_2$ + 2$H_2O$ → 4$NO_2^-$ + 4$H^+$.[42,43] The results of Fig. 3 are consistent with the results of Fig. 2. In addition, we may get the conclusion that $H_2O_2$ and $NO_2^-$ affect each other and their concentrations are suppressed by each other.

Plasma-activated DI water was applied to cancer cells and untreated DI water was used as control. Fig. 4 shows the cell viability of the human breast cancer cells and gastric cancer cells exposed to DI water, He plasma-activated DI water, Ar plasma-activated DI water, and $N_2$ plasma-activated DI water for 48 hours. The cell viability of MDA-MB-231, MCF-7, and NCI-N87 decreased by approximately 55%, 69%, and 65% compared with DI water when treated with He plasma-activated DI water at 30 min, respectively. The cell viability of three cancer cells exhibits a time-dependent behavior when treated by He and Ar plasma-activated DI water. For $N_2$ plasma-activated DI water, the cell viability of three cancer cells decreased firstly, followed by an increase as treatment time increased. The strongest effect can be observed in the case of He-feed plasma for MCF-87 cells at 30 min treatment. It's known that ROS can induce cell death by apoptosis and necrosis. Plasma can generate a series of ROS, including Atomic oxygen (O), superoxide ($O_2^-$), ozone ($O_3$), hydroxyl radical (•OH), singlet delta oxygen (SOD, $O_2(^1\Delta g)$), hydrogen peroxide ($H_2O_2$), and etc.[44,45] Several types of ROS has been proved useful on cancer cell death induction.[1,33] For example, $O_3$ leads to cell death via the formation of biologically active ROS and RNS in aqueous media,[46,47] $O_2(^1\Delta g)$ produces oxidative damage and selectively kills tumor cells,[48] and $O_2^-$ activates mitochondrial-mediated apoptosis.[49,50] In addition, aquaporins (AQPs) help to facilitate the passive diffusion of $H_2O_2$, resulting in cell death.[33] Overall, the above results suggest that ROS plays a major role in the interaction of plasma-activated DI water with three types of cancer cells.

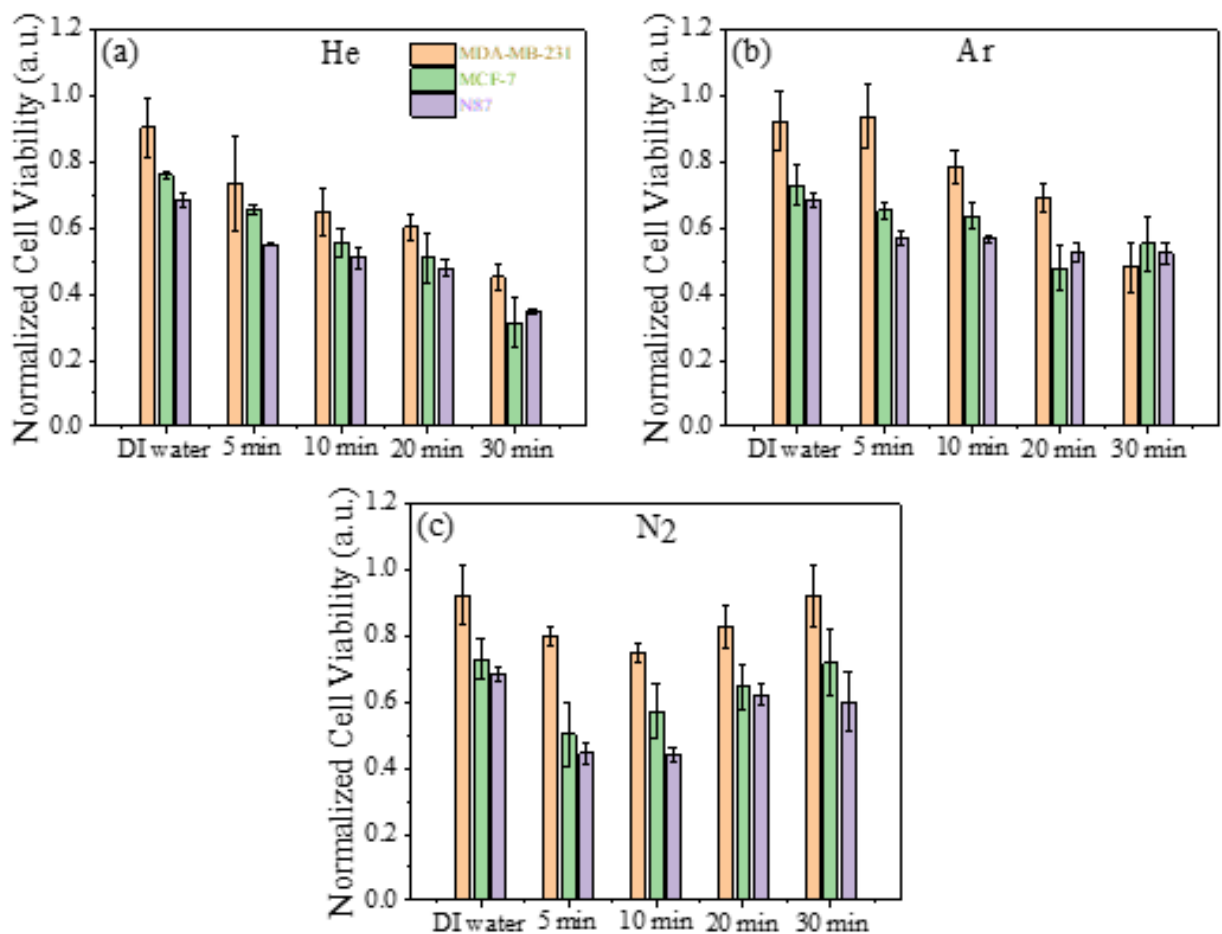

**Fig. 4**. The effects of the five solutions: DI water (0 min), and plasma-activated DI water using He (a), Ar (b), and $N_2$(c) during 5, 10, 20, and 30 min treatment. Cell relative metabolic activity of the human breast cancer cells (MDA-MB-231 and MCF-7) and human gastric cancer cells (NCI-N87) were measured at 48-hour incubation.

In conclusion, the CAP device under DI water was developed and the effect of plasma-activated DI water using Ar, He and $N_2$ as feeding gases on breast cancer cells (MDA-MB-231 and MCF-7) and gastric cancer cells (N87) were investigated. He plasma-activated DI water shows the highest concentration of ROS while N$_2$ plasma-activated DI water exhibits the highest concentration of RNS. ROS plays a major role in inducing cell death in MDA-MB-231, MCF-7, and N87 cancer cells. The He plasma-activated DI water during a 30 min treatment had the most significant effect in inducing MCF-7 cancer cell death.,